\documentclass{PoS}

\usepackage{amsmath}

\title{Towards Radiative Transitions in Charmonium}

\ShortTitle{Radiative Transitions in Charmonium}

\author{\speaker{Cian O'Hara}, Sin\'ead M. Ryan\\
        School of Mathematics, Trinity College\\
        Dublin 2, Ireland\\
        E-mail: \email{oharaci@tcd.ie, ryan@maths.tcd.ie}}

\author{{Graham Moir, Christopher E. Thomas}\\
        DAMTP, University of Cambridge,\\ 
        Centre for Mathematical Sciences, Wilberforce Road, Cambridge CB3 0WA, UK\\
        E-mail: \email{graham.moir@damtp.cam.ac.uk, C.E.Thomas@damtp.cam.ac.uk }\\

\\

\textup{(for the Hadron Spectrum Collaboration)}}

\abstract{We present preliminary calculations towards radiative transitions in charmonium using anisotropic $N_f = 2 + 1$ dynamical ensembles generated by the Hadron Spectrum Collaboration. With the use of newer technologies we aim to investigate transitions between states, including potential exotic charmonium states, lying higher in the spectrum than in previous studies. A crucial ingredient in this work is the use of variationally optimised interpolating operators which allow for a reliable determination of the three-point correlation functions needed. Using these operators, we perform first calculations of relevant three-point correlation functions before discussing future directions.}

\FullConference{34th annual International Symposium on Lattice Field Theory\\
		24-30 July 2016\\
		University of Southampton, UK \hfill \textnormal{DAMTP-2017-3}}

\begin{document}

\section{Introduction}

\noindent Lattice calculations of hadronic spectra have long been the definitive way of testing the validity of the theory of quantum chromodynamics in the non-perturbative regime. The comparison of high-precision spectral calculations with up to date experimental data allows for a rigorous test of the theory's description of the strong interaction. The accurate description of the energies of low lying states in the QCD spectrum has been an important benchmark of lattice studies for many years. Recently newer technologies and increased computational power have allowed for studies of higher-lying states and resonances. These studies allow for much more precise determinations than ever before, as well as providing valuable insight into previously unstudied regions such as hybrid or exotic states. In particular for charmonium and charmed mesons see Refs.~\cite{us,Liu,Graham}.

Charmonium, frequently characterised as the ``hydrogen atom'' of meson spectroscopy due to the fact that it is non-relativistic enough to be reasonably well described by certain potential models, is the perfect testing ground for a comparison of theory with experiment. Over the last decade there has been a renewed interest in spectral calculations due to the experimental discovery of many unexpected states. Specifically in the case of charmonium, so-called $X, Y, Z$ states highlight the need for a more complete theoretical understanding of the hadronic spectrum, be they hybrid mesons, tetra-quarks or some other hitherto unknown form of matter. Similarly, in the charm-light sector, states such as the $D_{s0}(2317)^{\pm}$ and $D_{s1}(2460)^{\pm}$ have been found to have much narrower widths than expected. For recent reviews of these topics see \cite{SwansonXYZ,Pren}.

To accurately understand the spectrum we must also investigate properties other than masses, such as decay rates. In the charmonium system the lowest-lying states lie below the $D\bar{D}$ threshold, resulting in relatively narrow widths due to the absence of OZI allowed strong decays. This means that radiative transitions, transitions from an initial state to final state via the emission of a photon, can have significant experimentally accessible branching ratios, and lattice calculations of such quantities can go a long way to provide valuable insight for experiment. 


As an example, an investigation of these transition rates, through calculations of electromagnetic vector current matrix elements, gives access to a hadron's photocoupling as well as it's underlying quark and charge distributions. Calculations of this photocoupling, a measure of how strongly the hadron couples to the photon, would be of particular interest to experimentalists as a test of the expectation that the photoproduction rate of hybrid states is large.

In these proceedings we will discuss the technologies to be used to probe these transitions on dynamical ensembles, first introduced in a recent calculation of radiative transitions in the light sector in Ref.~\cite{Shultz}, with the ultimate aim of studying transitions between excited states in the charmonium spectrum. There have been some investigations into lower lying transitions, such as those seen in references \cite{Becirevic2013,Becirevic2015,Chen,Donald}. However this will be the first dynamical calculation of excited charmonium transitions using the Hadron Spectrum Collaboration's approach. Previous studies of radiative transitions in charmonium on quenched lattices can be seen in Ref.~\cite{2009CharmTrans}. 

\section{Techonology}

\noindent We are interested in calculating quark-field vector current matrix elements between states $m$ and $n$ of the form $\langle m | j^{\mu} | n \rangle$ where $j^{\mu}$ is the standard vector current. These matrix elements encode, to leading order in $\alpha_{em}$, the coupling of mesons and baryons to the photon. For our analysis we have $j^{\mu}=\frac{2}{3}\bar{c}\gamma^{\mu}c$, where $\frac{2}{3}$ is the charm quark charge in units of $e$.

\begin{table}[t]
\begin{center}
\begin{tabular}{|c|c|c|c|c|}
\hline
 Lattice Volume & $M_{\pi}$ (MeV) & $N_{\rm cfgs}$ & $N_{tsrcs}$ & $N_{\rm vecs}$ \\ 
\hline
 $20^{3}\times 128$ & 391 & 50 & 1 & 128 \\
\hline
\end{tabular}
\caption{Information on the lattice gauge field ensemble used in this preliminary analysis, where $M_{\pi}$ is the pion mass, $N_{\rm cfgs}$ and $N_{\rm tsrcs}$ are the number of gauge field configurations and time-sources per configuration and $N_{\rm vecs}$ is the number of eigenvectors used in the distillation framework.}
\label{tab:lattice_details}
\end{center}
\end{table}

Due to constraints from Lorentz invariance, these matrix elements between a hadron $h$ of spin $J$ and helicity $\lambda$ along $\vec{p}$, and a second hadron $h'$ with $J',\lambda',\vec{p}'$ can be expanded as a sum over multiple form factors and Lorentz kinematic factors $K_i$, as discussed in Ref.~\cite{Shultz},

\begin{equation}
\langle h'_{J'}(\lambda',\vec{p}') | j^{\mu} | h_J(\lambda,\vec{p})\rangle = \sum_i K_i^{\mu}[h_{J'}'(\lambda',\vec{p}');h_J(\lambda,\vec{p})]F_i(Q^2).
\end{equation}

\noindent If the initial and final state hadrons are the same we extract the radiative form-factors $F_i(Q^2)$, where the photon's virtuality, $Q^2 = -q^2 = |\vec{p}'-\vec{p}|^2 - (E_{h'}(\vec{p}')-E_h(\vec{p}))^2$, measures the extent to which the photon is off shell. Transition form-factors are similarily extracted from matrix elements between different initial and final states. 

Technically, in the case of charmonium, we do not have radiative form-factors as hidden-charm mesons are eigenstates of charge-conjugation. The physical interpretation of this is that the photon couples equally to both the quark and anti-quark in the meson. In our analysis we choose to couple only to the quark, allowing us to probe the vector current's distribution as a function of virtuality within the meson. 


\subsection{Optimised operators and spectroscopy}

\noindent The key ingredient in any lattice spectroscopic calculation is the two-point function 

\begin{equation}
C_{ij}(t) = \langle 0|\mathbb{O}_i(t)\mathbb{O}_j^{\dagger}(0)|0\rangle.
\end{equation}

\noindent Using the distillation framework \cite{Dist}, we compute correlation functions for a large basis of operators $\mathbb{O}_i$, and diagonalize $C_{ij}(t)$ by solving the generalized eigenvalue problem (GEVP),

\begin{equation}
C_{ij}(t)v^{(n)} = \lambda_n(t,t_0)C_{ij}(t_0)v^{(n)},
\end{equation}

\noindent for a carefully chosen reference timeslice $t_0$. From this we extract the generalised eigenvalues $\lambda_n$ and the generalised eigenvectors $v^{(n)}$, which are related to spectroscopic quantities of interest. In the construction of these correlation functions all of the interpolators used are of the form

\begin{equation}
\mathbb{O}^{[J,P,|\lambda|]}_{\Lambda,\mu}(\vec{p}) = \sum_{\hat{\lambda}=\pm|\lambda|} S^{\tilde{\eta},\hat{\lambda}}_{\Lambda,\mu}\mathbb{O}^{JP,\hat{\lambda}}(\vec{p}). 
\end{equation}

\noindent These operators, with momentum $\vec{p}$, have been subduced into the appropriate irreducible representations, $\Lambda$, and row $\mu$, of the lattice symmetry group using the subductions coefficients $S^{\tilde{\eta},\hat{\lambda}}_{\Lambda,\mu}$, where $\tilde{\eta} \equiv P(-1)^J$. The helicity operators, $\mathbb{O}^{J,\lambda}$, with continuum spin $J$, parity $P$ and helicity $\lambda$ are of the form 

\begin{equation}
\mathbb{O}^{J,\lambda}(\vec{p}) = \sum_M D^{(J)*}_{M\lambda}(R)\mathcal{O}^{J,M}(\vec{p}).
\end{equation}


\noindent Here $D$ is a Wigner-$D$ matrix and $R$ is the (active) transformation that rotates $(0,0,|\vec{p}|)$ to $\vec{p}$. $\mathcal{O}^{J,M} \sim CGs(m_1,m_2,m_3,...) \sum_{\vec{x}} e^{i\vec{p}\cdot\vec{x}} \bar{\psi}(\vec{x},t)\Gamma_{m_1} \overleftrightarrow{D}_{m_2} \overleftrightarrow{D}_{m_3} ... \psi(\vec{x},t)$ is a fermion bilinear operator with spin $J$ and spin $z$-component $M$, formed by coupling together vector-like gamma matrices $\Gamma$ and gauge-covariant derivatives $\overleftrightarrow{D}$ with the appropriate Clebsch Gordan coefficients. $\psi(\vec{x},t)$ are the distillation smeared quark fields. For more information on these operator constructions, and a discussion on the zero momentum case, see Refs.~\cite{Helicity,Dudek:2010wm}.

 


\begin{figure}[t!]
\centering
\includegraphics[scale=0.44]{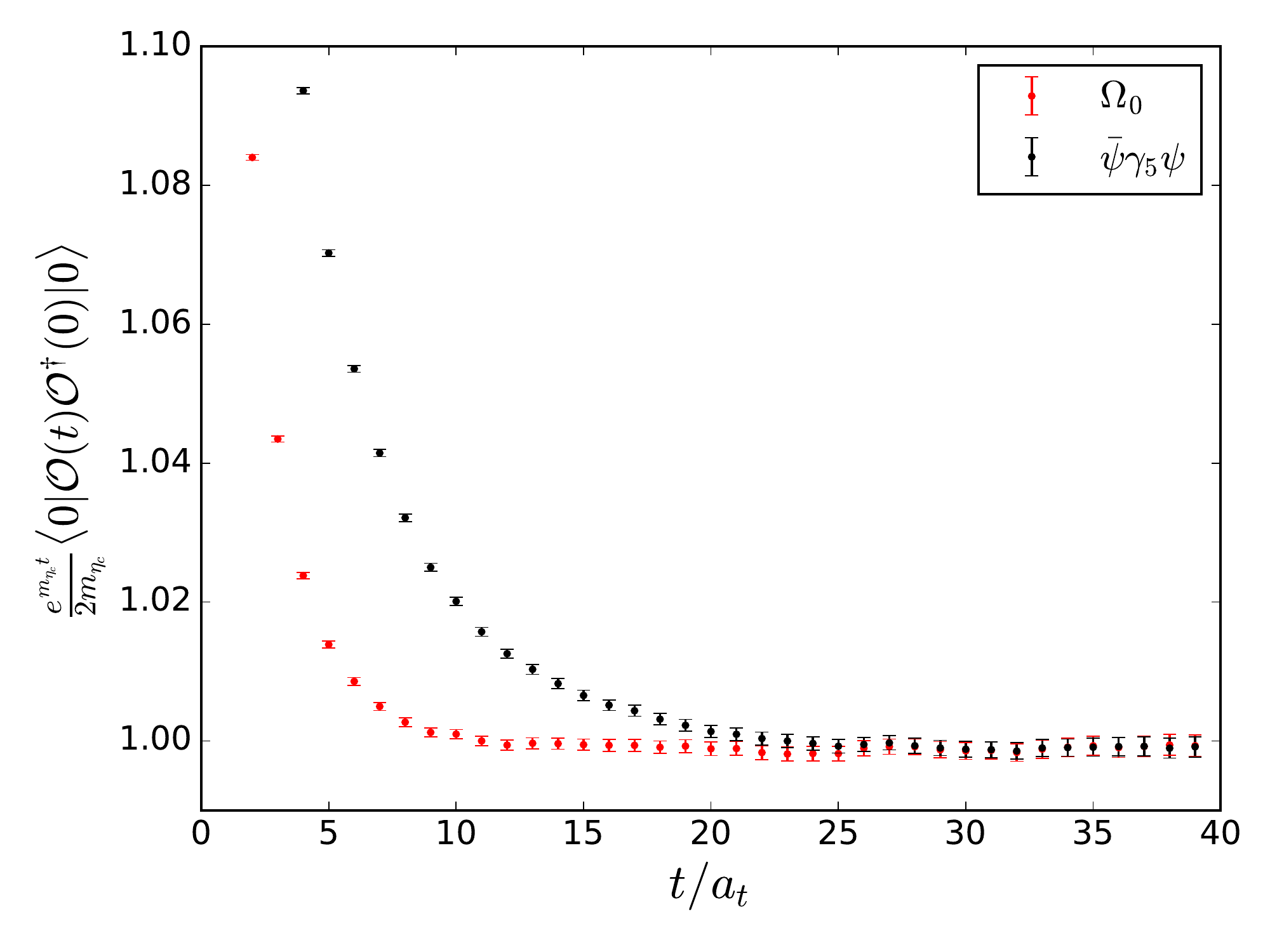}
\label{Impop}
\caption{$e^{m_{\eta_c}t}/2m_{\eta_c}\langle 0|\mathbb{O}(t)\mathbb{O}^{\dagger}(0)|0\rangle$ plotted for the $\eta_c$ at rest using the optimised $\eta_c$-like operator (shown in red) and the standard $\bar{\psi}\gamma_5\psi$ operator (in black). The earlier plateau in the case of optimised operators can be seen clearly.}
\end{figure}

In general, each of these operators will have some overlap with each state having the same quantum numbers. It is known that some linear combination of operators in the basis will overlap most strongly onto the desired eigenstates. It can be shown that the best estimates for the weights of this linear combination (in a variational sense) come from solving the GEVP, \cite{Dudek:2007}, motivating the formation of \emph{optimised operators} as

\begin{equation}
\label{eq:optimized_ops}
\Omega_{n,\Lambda,\mu}^{[J,P,|\lambda|]\dag}(\vec{p})= \sqrt{2E_n} e^{-E_nt_0/2}\sum_{i}v^{(n)}_i \mathbb{O}_{i,\Lambda,\mu}^{[J,P,|\lambda|]\dag}(\vec{p}),
\end{equation}

\noindent with $E_n$ being the energy of the $n^{th}$ eigenstate. For a demonstration of the feasability of using the GEVP method to extract multiple excited states in the charmonium spectrum, see Ref.~\cite{Liu}. 

Correlators utilising these optimised operators show a plateau a number of timeslices earlier than those using the unimproved equivalent, as seen in Figure 1. This allows for the reliable extraction of spectral information at earlier times, which is of special importance for transitions between excited states.


\subsection{Extracting form factors}


\noindent To access matrix elements of interest in the extraction of radiative form-factors one must look at three point correlation functions, with vector current insertion $j^{\mu}$

\begin{equation}
C^{\mu}_{ij}(\Delta t,t) = \langle 0|\mathbb{O}_i(\Delta t)j^{\mu}(t)\mathbb{O}_j^{\dagger}(0)|0\rangle.
\end{equation}

\noindent Here $\mathbb{O}_i,\mathbb{O}_j^{\dagger}$ are from the basis of interpolating fields and $j^{\mu}$ is inserted at time $t$ such that $ \Delta t >t >0$. All operators used, as well as the vector current insertion are projected onto definite momentum. As before, these interpolators will have some overlap with all states having the same quantum numbers. The correlation functions can be expanded as  
 
\begin{equation}
\small C^{\mu}_{ij}(\Delta t,t) = \sum_{m,n} \frac{1}{2E_m}\frac{1}{2E_n}e^{-E_m(\Delta t - t)} e^{-E_nt} \langle 0| \mathbb{O}_i(0) |m\rangle {\langle m |j^{\mu}(0)|n\rangle} \langle n | {\mathbb{O}}^{\dagger}_j(0)|0 \rangle.
\end{equation}

\noindent This sum contains contamination from many excited states as well as the desired matrix element, $\langle m | j^{\mu}(0) | n \rangle$. Using improved interpolators which project onto the states $m$ and $n$, and a normalization where $\langle n | \Omega_n^\dag|0\rangle= 2E_n$ (due to Eqn.~\ref{eq:optimized_ops}), the three point function simplifies to 

\begin{equation}
C^{\mu}_{mn}(\Delta t,t) = \langle 0| \Omega_m(\Delta t) j^{\mu}(t) \Omega^{\dagger}_n(0)|0 \rangle = e^{-E_m(\Delta t-t)}e^{-E_nt} \langle m|j^{\mu}(0)|n\rangle + ...
\end{equation}

\noindent The ellipsis represents left-over contamination from higher excited states that is suppressed with the use of appropriately chosen $t$ and $\Delta t$. The desired matrix elements are then extracted by dividing out the euclidean time dependence from the three point functions. For more information on the distillation method and the use of \emph{generalised perambulators} in the computation of three point functions see Ref.~\cite{Shultz}.

\subsection{Renomalisation and improvement of the vector current}

\begin{figure}[t!]
\centering
\includegraphics[scale=0.4]{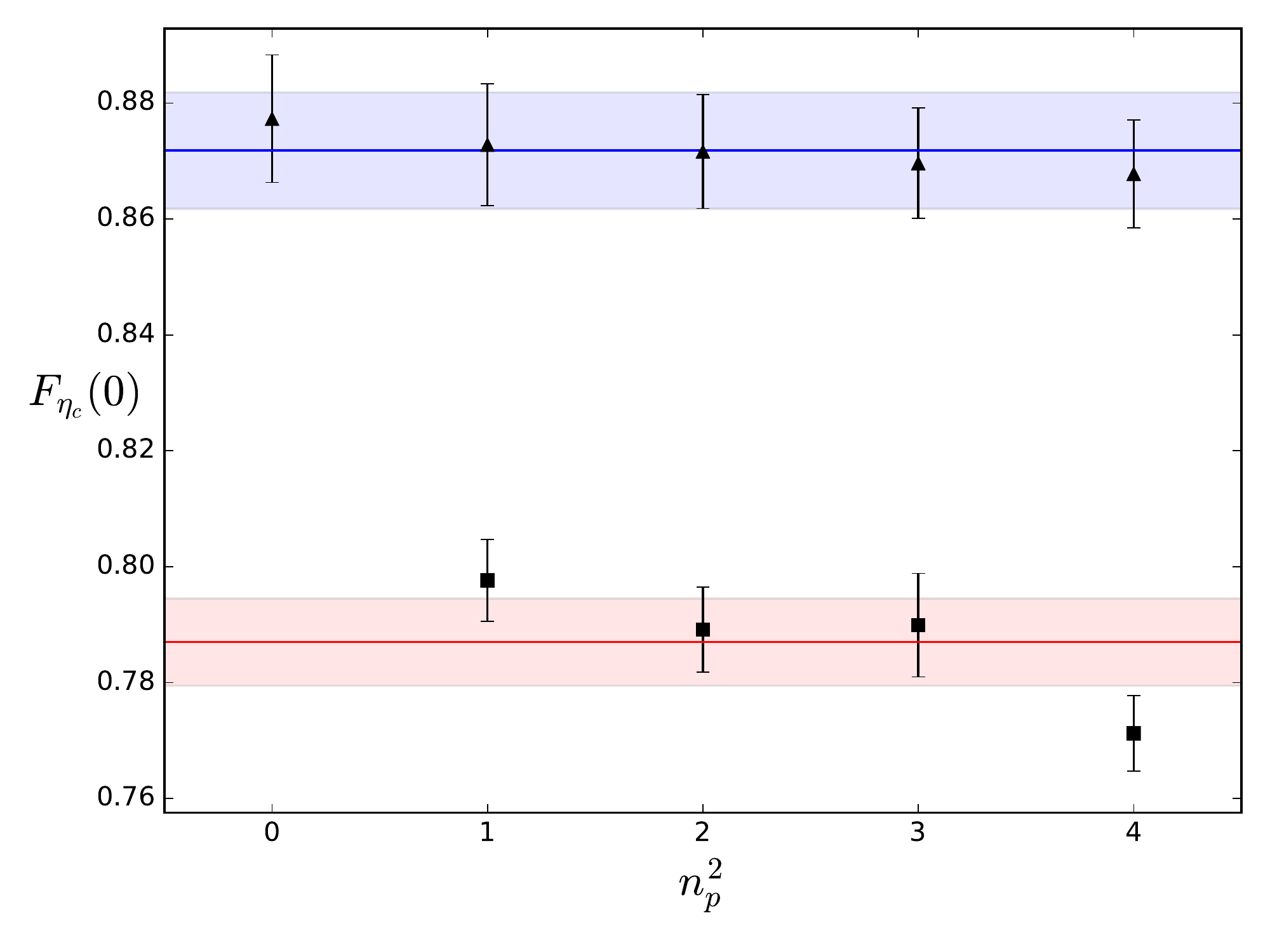}
\label{Z}
\caption{Preliminary value for zero momentum transfer form factor on 50 configurations. Plotted are the spatial (squares) and temporal (triangles) unrenormalised $\eta_c$ form factor for different values of $n_p^2 = (\frac{L}{2\pi})^2|\vec{p}|^2$, with the bands giving the statistical average and one sigma uncertainty.}
\end{figure}

\noindent The local vector current $\bar{\psi}\gamma^{\mu}\psi$, is not conserved on the lattice, and must be multiplicatively renormalised by a factor $Z_V$. This factor can be different for the spatial and temporal currents, $Z^s_V$ and $Z^t_V$ respectively, due to the anisotropic lattice. We choose to extract $Z_V$ from the pseudoscalar charge form-factor, $F_{\eta_c}^{lat}$, which appears in the decomposition of correlation functions with improved $\eta_c$ interpolators at the source and sink for zero momentum transfer. We insist that this should take its continuum value of unity, such that

\begin{equation}
Z_V = \frac{F_{\eta_c}^{cont.}(0)}{F_{\eta_c}^{lat.}(0)} = \frac{1}{F_{\eta_c}^{lat.}(0)}.
\end{equation}

\noindent Using improved $\eta_c$ operators in the definition above we find

\begin{equation}
C_{\eta_c \eta_c}^{\mu}(\Delta t,t) = e^{-E_{\eta_c}\Delta t}\langle \eta_c| j^{\mu} | \eta_c \rangle,
\end{equation}
 
\noindent up to some small pollution from higher lying states which should be minimal. This can be checked to be insignificant by varying $\Delta t$. In the case of the $\eta_c$, the decomposition of the matrix element into form factors is given as 

\begin{equation}
\langle \eta_c| j^{\mu} | \eta_c \rangle = (p + p')^{\mu}F_{\eta_c}(Q^2).
\label{etacdecomp}
\end{equation}


\noindent As an illustrative example, when both source and sink particles have the same momenta we find, where there is no sum over $\mu$,

%
%

\begin{equation}
F_{\eta_c}(0) = \frac{1}{2p^{\mu}} \langle \eta_c| j^{\mu} | \eta_c \rangle =  \frac{1}{2p^{\mu}}e^{E_{\eta_c}\Delta t}C^{\mu}_{\eta_c \eta_c}(\Delta t,t).
\end{equation}


\noindent  Figure 2 shows a preliminary extraction for one timesource on 50 of the available configurations of the unrenormalised zero momentum transfer $\eta_c$ form factor. The difference between the temporal and spatial values extracted for $F_{\eta_c}$ highlights the need for a different renormalisation factor $Z_V$ for the temporal and spatial directions.



\section{Future directions}

\noindent A possible enhancement would be to use an improved current in future calculations. The anisotropic discretisation introduces a tree level $O(a)$ improvement term which amounts to replacing the Euclidean current $j_{\mu} = \bar{\psi}\gamma_{\mu}\psi$ with 

\begin{eqnarray}
j_0 &=& Z^t_V(\bar{\psi}\gamma_0\psi + \frac{1}{4}\frac{\nu_s}{\xi}(1-\xi)a_s\partial_j(\bar{\psi}\sigma_{0j}\psi)) \nonumber\\
j_k &= &Z^s_V(\bar{\psi}\gamma_k\psi + \frac{1}{4}(1-\xi)a_t\partial_0(\bar{\psi}\sigma_{0k}\psi)),
\end{eqnarray}

\noindent where $\xi = a_s/a_t$ is the anisotropy and $\nu_s$ is a parameter appearing in the anisotropic fermion action. Note that a mass-dependent term has been absorbed into our renormalisation factors. 
As $\sigma_{\mu\nu} = \frac{i}{2}[\gamma_{\mu},\gamma_{\nu}]$, improving the current amounts to in addition calculating correlators of the form $\langle...|\bar{\psi}\gamma_0\gamma_i\psi|...\rangle$

In summary, we have introduced the technology needed to compute radiative transition form-factors for transitions involving excited states in the charmonium spectrum. The introduction of optimised operators which overlap strongly with a single state in the spectrum was motivated by the need to study transitions between higher excited states in the spectrum, which have been relatively unstudied up until now and will encompass the next stage of this work.

\section*{Acknowledgements}

We thank our colleagues within the Hadron Spectrum Collaboration. COH acknowledges support from the School of Mathematics at Trinity College Dublin. GM acknowledges support from the Herchel Smith Fund at the University of Cambridge. SMR acknowledges support from Science Foundation Ireland [RFP-PHY-3201]. CET acknowledges support from the U.K. Science and Technology Facilities Council (STFC) [grant ST/L000385/1].


\begin{thebibliography}{99}

\bibitem{us}
G.K.C. Cheung, C. O'Hara, G. Moir, M. Peardon, S.M. Ryan, C.E. Thomas, D. Tims,
\textit{Excited and exotic charmonium, $D_s$ and $D$ meson spectra for two light quark masses from lattice QCD},
\emph{JHEP} {\bf 12} 089 (2016), arXiv:1610.01073.

\bibitem{Liu}
L. Liu \textit{et al}.
\textit{Excited and exotic charmonium spectroscopy form lattice QCD}
\emph{JHEP} {\bf 07} 126 (2012), arXiv:1204.5425.

\bibitem{Graham}
G. Moir, M. Peardon, S.M. Ryan, C.E. Thomas, L. Liu,
\textit{Excited spectroscopy of charmed mesons from lattice QCD}
\emph{JHEP} {\bf 05} 021 (2013), arXiv:1301.7670


\bibitem{SwansonXYZ}
E.S. Swanson,
\textit{XYZ States: Theory Overview,}
\emph{AIP Conf. Proc.} {\bf 1735} (2016) 020013, arXiv:1512.04853.

\bibitem{Pren}
E. Prencipe,
\textit{Hadrons with c-s content: past, present and future}
\emph{Proc. 53rd International Winter meeting on Nuclear Physics (Bormio 2015); Bormio, Italy, January 26-30,2015},
arXiv:1510.03053.


\bibitem{Shultz}
C.J. Shultz, J.J. Dudek, R.G. Edwards,
\textit{Excited meson radiative transitions from lattice QCD using variationally optimized operators,}
\emph{Phys.Rev.} 11 {\bf D91} 114501 (2015), arXiv:1501.07457. 

\bibitem{Becirevic2013}
D. Becirevic, F. Sanfilippo,
\textit{Lattice QCD study of the radiative decays $J/\psi\to\eta_c\gamma$ and $h_c\to \eta_c\gamma$,}
\emph{JHEP} {\bf 01} 028 (2013), arXiv:1206.1445.

\bibitem{Becirevic2015}
D. Becirevic, M. Kruse, F. Sanfilippo,
\textit{Lattice QCD estimate of the $\eta_c (2S) \rightarrow J/\psi \gamma$ decay rate,}
\emph{JHEP} {\bf 05} 014 (2015), arXiv:1411.6426

\bibitem{Chen}
Y. Chen \textit{et al}.
\textit{Radiative transitions in charmonium from $N_f=2$ twisted mass lattice QCD,}
\emph{Phys. Rev.} {\bf D84} 034503 (2011), arXiv:1104.2655.

\bibitem{Donald}
G.C. Donald \textit{et al}.
\textit{Precision tests of the $J/{\psi}$ from full lattice QCD: mass, leptonic width and radiative decay to ${\eta}_c$,}
\emph{Phys. Rev.} {\bf D86} 094501 (2012), arXiv:1208.2855

\bibitem{2009CharmTrans}     
J.J. Dudek, R.G. Edwards, C.E. Thomas,
\textit{Exotic and excited-state radiative transitions in charmonium from lattice QCD,}
\emph{Phys.Rev.} {\bf D79} 094504 (2009), arXiv:0902.2241.

\bibitem{Dist}
Hadron Spectrum Collaboration, M.J. Peardon et al. \textit{A novel quark-field creation operator construction for hadronic physics in lattice QCD}
\emph{Phys. Rev.} {\bf D80} (2009) 054506, arXiv:0905.2160.

\bibitem{Helicity}
C.E. Thomas, R.G. Edwards, and J.J. Dudek,
\textit{Helicity operators for mesons in flight on the lattice,}
\emph{Phys. Rev.} {\bf D85} 014507 (2012), arXiv:1107.1930.

\bibitem{Dudek:2010wm}
J.J Dudek, R.G. Edwards, M.J. Peardon, D.G. Richards and C.E. Thomas,
\textit{Toward the excited meson spectrum of dynamical QCD,}
\emph{Phys. Rev.} {\bf D82} 034508 (2010), arXiv:1004.4930.


\bibitem{Dudek:2007}
J.J. Dudek, R.G. Edwards, N. Mathur, D.G. Richards, 
\textit{Charmonium excited state spectrum in lattice QCD,}
\emph{Phys. Rev.} {\bf D77} 034501 (2008), arXiv:0707.4162.


\end{thebibliography}
\end{document}